\documentclass[aps,preprint,showpacs,groupedaddress]{revtex4-1}
\usepackage{amssymb}
\usepackage{mathrsfs}
\usepackage{pstricks}
\usepackage{graphicx}
\usepackage{hyphenat}
\usepackage{hyperref}

\begin{document}


\title{\bf NLO critical exponents of O($N$) lambda $\phi^{4}$ scalar field theories in curved spacetime}


\author{H. A. S. Costa}
\email{helderfisica@gmail.com}
\author{P. R. S. Carvalho}
\email{prscarvalho@ufpi.edu.br}
\affiliation{\it Departamento de F\'\i sica, Universidade Federal do Piau\'\i, 64049-550, Teresina, PI, Brazil}




\begin{abstract}
In this paper we investigate analytically the conformal symmetry influence on the next-to-leading order radiative quantum corrections to critical exponents for massless O($N$) $\lambda\phi^{4}$ scalar field theories in curved spacetime. We renormalize the theory by applying the BPHZ method. We find that the critical exponents are the same as that of flat spacetime, at least at the loop order considered. We argue that this result agrees perfectly with the universality hypothesis.
\end{abstract}


\maketitle


\section{Introduction}\label{Introduction} 

\par In general, physical phenomena are not affected by the state of their observers. For example, the special theory of relativity was built by demanding that an experimentally measured physical quantity has to display the same result when measured by an observer at rest or in relative motion to the former with constant velocity. From this requirement, it has emerged the concept of symmetry transformation of observers, \emph{i.e.} a given physical quantity will present the same value if measured by any of the referred observers. In the aforementioned case, the symmetry transformation is the Lorentz symmetry. Others symmetries were discovered along the past years as continuous (time translation, spatial translation, spatial rotation, gauge U($1$), gauge SU($2$), gauge SU($3$) etc) and discrete (time reversal, spatial inversion, charge change etc) ones \cite{H.Georgi,Greiner,G.CostaandG.Fogli} for citing just a few of them. Many of these symmetries are needed for a complete understanding of the behavior of physical systems at the high energy physics realm \cite{Itzykson,Peskin,Kleinertt}. On the other hand, at the condensed matter physics scenario, the symmetry physical implications have also turned out to be profound \cite{PhysRevLett.28.240,PhysRevLett.28.548,Wilson197475,PhysRevD.7.2911,PhysRevB.4.3174,PhysRevB.4.3184,PhysRevD.2.1478,PhysRevD.6.419}. In fact, in the phase transitions and critical phenomena domain, the universal critical behavior of a given system undergoing a continuous phase transition is characterized by a corresponding set of six universal critical exponents \cite{Stanley}. Fortunately, many completely distinct physical systems as a fluid and a ferromagnet can have their phase transition behaviors characterized by the same set of critical exponents. For that, they have to have in common their dimension $d$, $N$ and symmetry of some $N$-component order parameter (for example, the order parameter can be the magnetization when we are leading with magnetic systems) if the interactions are of short- or long-range type. When this happens, we say that these distinct physical systems belong to the same universality class. We will be concerned in this work with the O($N$) one. It includes the specific models: Ising ($N = 1$), XY ($N = 2$), Heisenberg ($N = 3$), self-avoiding random walk ($N = 0$), spherical ($N \rightarrow \infty$) etc. \cite{Pelissetto2002549}. The influence of the $d$ and $N$ parameters on the critical exponents values is easier to probe \cite{PhysRevB.86.155112,PhysRevE.71.046112,PhysRevLett.110.141601,Butti2005527,PhysRevB.54.7177} than that of symmetry of course. Some works were purposed to investigate the result of considering the symmetry effect on the critical exponents values in flat spacetime \cite{PhysRevE.78.061124,Trugenberger2005509,CARVALHO2017290,Carvalho2017}. We now probe the conformal symmetry effect in curved spacetime. 

\par In this paper we compute analytically the next-to-leading order (NLO) critical exponents for massless O($N$) $\lambda\phi^{4}$ scalar field theories in curved spacetime. In this case, massless theories represent systems at the critical temperature $T_{c}$, since the mass of the fluctuating quantum field corresponds to $T - T_{c}$, where $T$ some arbitrary temperature and $T_{c}$ is the critical one. For that, we employ the field-theoretic renormalization group approach based on the dimensional regularization of Feynman diagrams in $\epsilon$-expansion ($\epsilon = 4 - d$), where the theory is renormalized through the massless Bogoliubov-Parasyuk-Hepp- \\ Zimmermann (BPHZ) method \cite{BogoliubovParasyuk,Hepp,Zimmermann}. For attaining that task, we have to obtain the loop corrections to critical exponents, at least, at the finite two-loop order. As the one-loop contribution to field renormalization is vanishing, for obtaining the renormalized field at next-to-leading level, we have to consider three-loop diagrams. The field-theoretic renormalization group approach for evaluating critical exponents \cite{Amit,ZinnJustin} is a formulation in which the information about the critical properties of the system is contained in the critical properties of the $1$PI vertex parts with their amputated external legs \cite{Kleinert}. They are defined as ensemble averages of fluctuating quantum fields $\phi$ with weight probability density given by the Lagrangian density of the system. These fluctuating quantum fields are such that, their mean values are identified to the magnetization when we treat magnetic systems, for example. The $1$PI vertex parts are initially divergent. These divergences can be absorbed into constants called renormalization constants. These constants contain the divergent structure of the theory and are used for evaluating the $\beta$-function, field and composite field anomalous dimensions. When these anomalous dimensions are computed at the nontrivial value in which $\beta = 0$, we obtain the radiative quantum corrections to the critical exponents. As there are four relations among the critical exponents, the scaling relations, we must evaluate just two of them independently \cite{Amit}. If we want a precise determination of the critical exponents, we have to consider the fluctuations of the quantum field. This is approached by the renormalization group technique. In fact, this mathematical tool was designed to be capable of taking into account the nontrivial interaction among the many degrees of freedom at all scales. Then, we can develop a perturbative theory in which how much more perturbative orders are attained, more precise results are obtained. If we do not consider that quantum fluctuations, we do not obtain satisfactory results. This situation is called the mean-field or Landau approximation for the critical exponents. As the Lagrangian density is essential for obtaining the $1$PI vertex parts, it has to be defined from the very beginning. From a general Lagrangian density, initially containing any powers of $\phi$ \cite{PhysRevD.49.6767}, we have that only even powers of $\phi$ must survive because they have to preserve the $\phi \rightarrow -\phi$ symmetry which represents the known symmetry under the simultaneous change of all spin orientations in magnetic systems for example \cite{Stanley}. The other constraint which turns possible eliminate many terms among the allowed ones is that only terms with canonical dimension smaller or equal to four will remain. They are called relevant or marginal operators and give contributions to the critical properties of the system, while the ones with canonical dimensions greater than four are called irrelevant and can be discarded \cite{Amit}. Now, for introducing the coupling between the fluctuating field and the curved background, we have to show that the only allowed interaction needed to describe the critical properties of the system will be the non-minimal one $\xi R\phi^{2}$. The parameter of the non-minimal interaction is $\xi$ and $R$ is the scalar curvature $R = g^{\mu\nu}R_{\mu\nu}$. This interaction is allowed and we justify why: namely, such a term is required to provide UV renormalizability starting from one loop order and beyond one loop the conformal value $1/6$ is not a fixed point \cite{PhysRevD.25.1019,Buchbinder.Odintsov.Shapiro,Parker.Toms,0264-9381-25-10-103001}. Instead of defining the fluctuating quantum field on a flat spacetime, it is now defined on a Rimmannian manifold. Thus the Lagrangian must be invariant under conformal transformations and integrated out over a covariant volume for obtaining the respective action of the theory. So we have to incorporate the term $\sqrt{g}$ to the theory, where $g$ is given by $g = det(g^{\mu\nu})$. Thus the Lagrangian density considered in this paper will be proportional to that of flat spacetime and containing the non-minimal interaction. In the massless theory approached here there are some peculiarities. In general, the divergences present in massless theories lie at small momentum scales and they are called infrared divergences. In the present theory, we can not get rid these divergences unless we set $\xi = 1/6$ at $d = 4$ \cite{0305-4470-13-2-023} which leads to a renormalizable theory with conformal invariance. Another feature which is a nontrivial one and a consequence of the renormalizability of the theory, \emph{i. e.} the theory is satisfactory only when $\xi = 1/6$, is that its one-loop contribution is vanishing. Thus we have to attain, at least, the next-to-leading level for obtaining a nonvanishing radiative quantum correction for this parameter, analogously to the field renormalization endeavor whose one-loop contribution is vanishing.     

\par In this paper we renormalize the theory at next-to-leading order in Sec. \ref{Next-to-leading order renormalization}. In Sec. \ref{Critical exponents up to next-to-leading order} we evaluate the critical exponents up to NLO. We conclude this work in Sec. \ref{Conclusions} with our final considerations.

\section{Next-to-leading order renormalization}\label{Next-to-leading order renormalization}

\par The renormalization scheme used in this paper to renormalize the theory is the BPHZ one \cite{BogoliubovParasyuk,Hepp,Zimmermann}. In this method, we start from a given loop order in which the Feynman diagrams of the $1$PI vertex parts are divergent at that order. Then we add terms to the diagrammatic expansions for the $1$PI vertex parts such that new finite ones are obtained. These added terms are called counterterm diagrams. The new finite $1$PI vertex parts can be now viewed as been generated by a new renormalized Lagrangian density containing new terms which in turn generate the counterterms diagrams. Thus, we obtain a finite Lagrangian density representing a finite theory at that order. We can proceed to the next order in which the theory is divergent and apply the same procedure for obtaining another Lagrangian density generating others counterterm diagrams and so on up to any desired order in perturbation theory such that we always attain a renormalized theory. Then we can always start from the renormalized theory at a given loop level. Thus we can start from the renormalized Lagrangian density  
\begin{eqnarray}\label{bare Lagrangian density}
&&\mathcal{L} = \sqrt{g}\frac{1}{2}\left(\partial_{\mu}\phi\partial^{\mu}\phi + \xi R\phi^{2} +  t\phi^{2}\right) +  \sqrt{g}\frac{\mu^{\epsilon}f}{4!}\phi^{4},
\end{eqnarray}
where it is defined on a $d$-dimensional curved background with Rimannian metric signature and $\phi$, $\lambda$, $t$ and $\xi$ are the renormalized field, coupling constant, composite field coupling constant and non-minimal interaction parameter, respectively. They are related to their bare counterparts through
\begin{eqnarray}\label{huytr}
\phi = Z_{\phi}^{-1/2}\phi_{0}, 
\end{eqnarray} 
\begin{eqnarray}\label{huytrooiuy}
f = \mu^{-\epsilon}\frac{Z_{\phi}^{2}}{Z_{f}}f_{0}, 
\end{eqnarray} 
\begin{eqnarray}\label{huytroo}
t = \frac{Z_{\phi}}{Z_{\phi^{2}}}t_{0}
\end{eqnarray} 
and
\begin{eqnarray}
\xi = \frac{Z_{\phi}}{Z_{\xi}}\xi_{0},
\end{eqnarray}
where $f$ is the dimensionless renormalized coupling constant and $\mu$ some arbitrary momentum scale parameter. From all $1$PI vertex parts, we have to consider only just a few of them, namely the primitively divergent $\Gamma^{(2)}$, $\Gamma^{(4)}$ and $\Gamma^{(2,1)}$ ones, since the higher $1$PI vertex parts can be obtained from the primitively divergent ones through the skeleton expansion \cite{ZinnJustin}. So, if we renormalize the primitively divergent vertex parts, the higher ones turn out to be automatically renormalized. In general, the critical exponent $\nu$ is computed in a massive theory through the renormalization of a mass term. The same task in a massles theory is attained through the renormalization of some operator called composite operator $\phi^{2}\equiv\phi(y)\phi(y)$. It is not simply the square of the field but it is composed of two fields evaluated at the same point of spacetime. This operator generates the bare composite $1$PI $\Gamma_{0}^{(2,1)}$ vertex part, to be renormalized, perturbatively in terms of $t_{0}$. Although the mass term and the $t$ one are distinct, both furnish the necessary information to evaluate the critical exponent $\nu$ \cite{ZinnJustin}. The loop expansion for the renormalized $1$PI vertex parts, up to NLO is given by \cite{Amit}
\begin{eqnarray}\label{fdhjdii}
\Gamma^{(2)} = \parbox{12mm}{\includegraphics[scale=1.0]{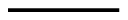}}^{-1}  - \frac{1}{6}\hspace{1mm}\parbox{12mm}{\includegraphics[scale=1.0]{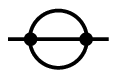}}  -  \frac{1}{4}\hspace{1mm}\parbox{10mm}{\includegraphics[scale=0.8]{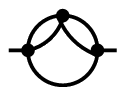}} - \frac{1}{3} \mathcal{K}
  \left(\parbox{10mm}{\includegraphics[scale=1.0]{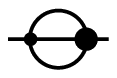}} \right), 
\end{eqnarray}
\begin{eqnarray}\label{fdijgihu}
&&\Gamma^{(4)} = - \hspace{1mm}\parbox{10mm}{\includegraphics[scale=0.09]{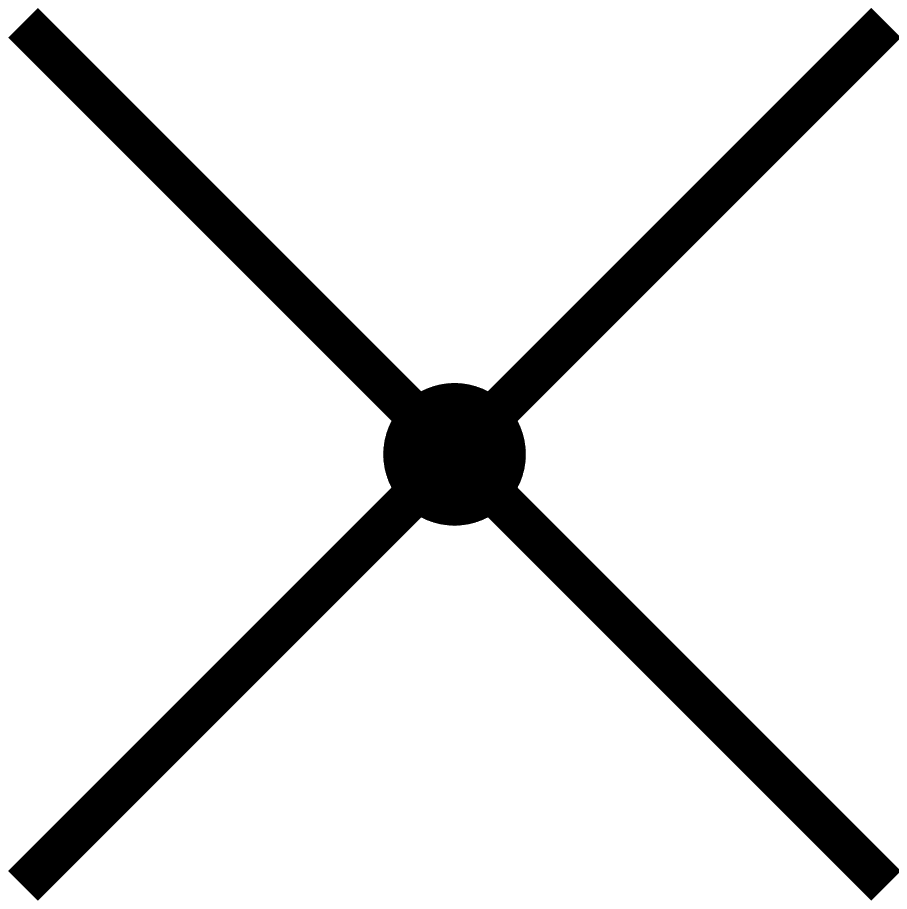}}  -  \frac{1}{2}\hspace{1mm}\parbox{10mm}{\includegraphics[scale=1.0]{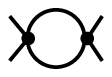}} + 2 \hspace{1mm} perm. - \frac{1}{4}\hspace{1mm}\parbox{16mm}{\includegraphics[scale=1.0]{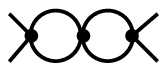}} + 2 \hspace{1mm} perm. -  \frac{1}{2}\hspace{1mm}\parbox{12mm}{\includegraphics[scale=0.8]{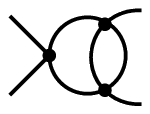}} + 5 \hspace{1mm} perm. - \nonumber \\ &&  \mathcal{K}
  \left(\parbox{10mm}{\includegraphics[scale=1.0]{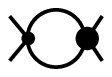}} + 2 \hspace{1mm} perm. \right),
\end{eqnarray}
\begin{eqnarray}\label{gtfrdesuuji}
\Gamma^{(2,1)} =  1 - \frac{1}{2}\hspace{1mm}\parbox{14mm}{\includegraphics[scale=1.0]{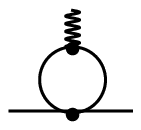}}  - \frac{1}{4}\hspace{1mm}\parbox{12mm}{\includegraphics[scale=1.0]{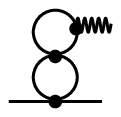}}  -  \frac{1}{2}\hspace{1mm}\parbox{12mm}{\includegraphics[scale=0.8]{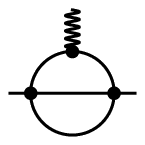}} - \frac{1}{2} \mathcal{K}
  \left(\parbox{12mm}{\includegraphics[scale=.2]{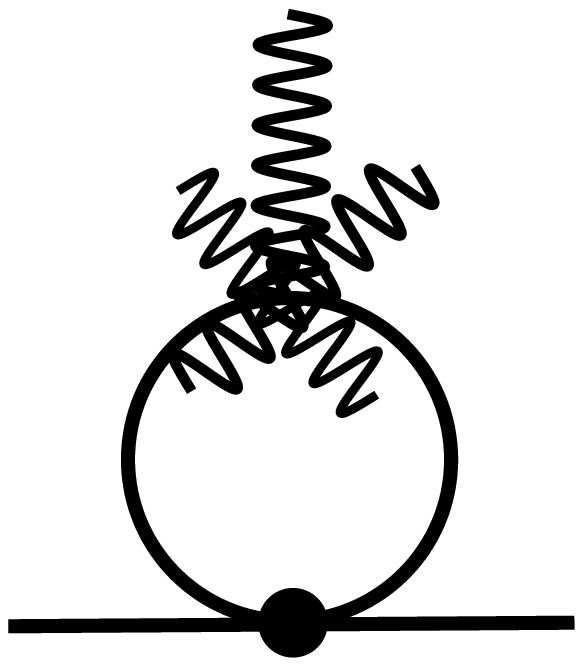}} \right) -  \frac{1}{2} \mathcal{K}
  \left(\parbox{12mm}{\includegraphics[scale=.2]{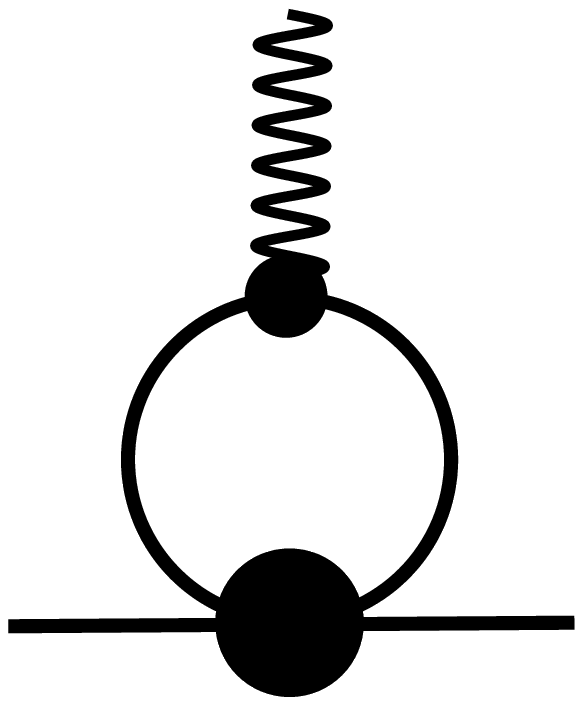}} \right)
\end{eqnarray}
where the internal line $\parbox{12mm}{\includegraphics[scale=1.0]{fig9.eps}}$ represents the classical propagator of scalar field $G_{0}(q)$ expanded in normal coordinates given by \cite{PhysRevD.20.2499}
\begin{eqnarray}\label{hdfughufghf}
G_{0}(q) = \frac{1}{q^{2}} + \frac{(1/3 - \xi)R}{(q^{2})^{2}} - \frac{2R_{\mu\nu}q^{\mu}q^{\nu}}{3(q^{2})^{3}} 
\end{eqnarray}
and 
\begin{eqnarray}
&&\parbox{6mm}{\includegraphics[scale=.1]{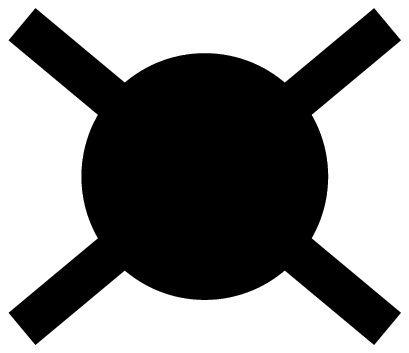}} = -\mu^{\epsilon}fc_{f}^{1} = -\frac{3}{2} \mathcal{K} 
\left(\parbox{10mm}{\includegraphics[scale=1.0]{fig10.eps}} \right),
\end{eqnarray}
\begin{eqnarray}
&&\parbox{16mm}{\includegraphics[scale=.2]{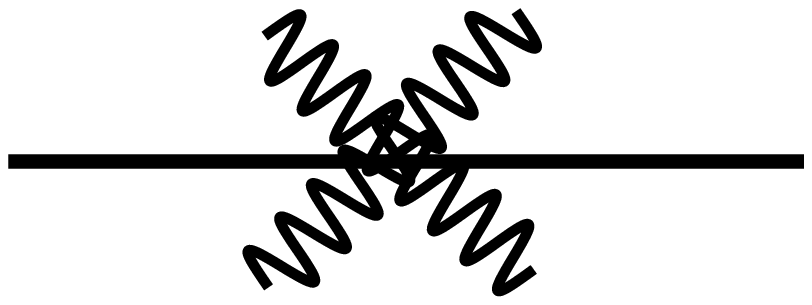}} = -c_{\phi^{2}}^{1}= -\frac{1}{2} \mathcal{K} 
\left(\parbox{11mm}{\includegraphics[scale=.8]{fig14.eps}} \right)
\end{eqnarray}
are the counterterms at one-loop order. We have expanded $G_{0}(q)$ only up to linear terms in $R$ and $R_{\mu\nu}$. Now we proceed to compute the critical exponents up to next-to-leading order.

\section{Critical exponents up to next-to-leading order}\label{Critical exponents up to next-to-leading order}

\par For computing the critical exponents up to NLO level, we have to evaluate the field and composite field anomalous dimensions evaluated at the nontrivial fixed point. For that, we have to compute the renormalization constants for field, composite field and coupling constant. They are related through the relations 
\begin{eqnarray}\label{kjjffxdzs}
\beta(f) = \mu\frac{\partial f}{\partial \mu},
\end{eqnarray}
\begin{eqnarray}\label{koiuhygtf}
\gamma_{\phi}(f) = \mu\frac{\partial\ln Z_{\phi}}{\partial \mu},
\end{eqnarray}
\begin{eqnarray}\label{qesdgdgd}
\gamma_{\phi^{2}}(f) = -\mu\frac{\partial\ln Z_{\phi^{2}}}{\partial \mu}.
\end{eqnarray} 
The $\beta$-function, field and composite field anomalous dimensions form the Callan-Symanzik equation
\begin{eqnarray}
\left[\mu\frac{\partial}{\partial\mu} + \beta(f)\,\frac{\partial}{\partial f} - \frac{1}{2}n\,\gamma_{\phi}(f) + l\,\gamma_{\phi^{2}}(f) + \beta_{\xi}(f)\,\xi\frac{\partial}{\partial \xi}\right] \Gamma^{(n)}(P_{1},...,P_{n};f,\mu) = 0,\quad
\end{eqnarray}
where the non-minimal coupling constant anomalous dimension $\gamma_{\xi}$ is defined by \cite{Buchbinder.Odintsov.Shapiro}
\begin{eqnarray}
\beta_{\xi}(f) = -\mu\frac{\partial \ln Z_{\xi}}{\partial\mu}.
\end{eqnarray}
We have analytically computed the diagrams of two-point function of Eq. (\ref{fdhjdii}). The respective results are displayed in the Appendix \ref{Evaluated three-loop diagrams}. We observe that in Eqs. (\ref{gjhkjhih}), (\ref{dfhuughfbjkdf}) and (\ref{djigjfjgi}) the terms proportional to the external momenta $P$ in quadratic form $P^{2}$ are the same as their corresponding flat spacetime values which in turn will contribute to field anomalous dimension evaluation as it is known \cite{Amit}
. Thus, the corresponding anomalous dimension in curved spacetime is the same as its flat spacetime counterpart  \cite{Amit}
\begin{eqnarray}\label{jkjkpfgjrftj}
\gamma_{\phi}(f) = \frac{N + 2}{72}f^{2} - \frac{(N + 2)(N + 8)}{1728}f^{3}.
\end{eqnarray}
The remaining diagrams, that of curved spacetime four-point and composite field functions of Eqs. (\ref{fdijgihu})-(\ref{gtfrdesuuji}) exhibit curved spacetime-dependent divergences in terms of $R$ and $R_{\mu\nu}$. They cancel out in the middle of calculations for the $\beta$-function and composite field anomalous dimension, at least at NLO. Thus we obtain the same flat spacetime expressions for the corresponding functions 
\begin{eqnarray}\label{reewriretjgjk}
\beta(f) = -\epsilon f + \frac{N + 8}{6}f^{2} - \frac{3N + 14}{12}f^{3},
\end{eqnarray} 
\begin{eqnarray}\label{gfydsguyfsdgufa}
\gamma_{\phi^{2}}(f) = \frac{N + 2}{6}f - \frac{5(N + 2)}{72}f^{2}.
\end{eqnarray} 
Similarly, the anomalous dimension $\gamma_{\xi}(u)$ results
\begin{eqnarray}\label{gfydsguyfsdgufa}
\beta_{\xi}(f) = \frac{(N + 2)}{216}f^{2}.
\end{eqnarray}
Thus, the NLO level critical exponents at curved spacetime obtained at the nontrivial fixed point \cite{Amit} 
\begin{eqnarray}\label{yagyaguhd}
f^{*} = \frac{6\epsilon}{(N + 8)}\left\{ 1 + \epsilon\left[ \frac{3(3N + 14)}{(N + 8)^{2}} \right]\right\}
\end{eqnarray}
through the relations $\eta\equiv\gamma_{\phi}(f^{*})$, $\nu^{-1}\equiv 2 - \gamma_{\phi^{2}}(f^{*})$ have their values identical to the corresponding flat spacetime ones \cite{Wilson197475}  
\begin{eqnarray}\label{eta}
\eta = \frac{(N + 2)\epsilon^{2}}{2(N + 8)^{2}}\left\{ 1 + \epsilon\left[ \frac{6(3N + 14)}{(N + 8)^{2}} -\frac{1}{4} \right]\right\},
\end{eqnarray}
\begin{eqnarray}\label{nu}
\nu = \frac{1}{2} + \frac{(N + 2)\epsilon}{4(N + 8)} +  \frac{(N + 2)(N^{2} + 23N + 60)\epsilon^{2}}{8(N + 8)^{3}}.\quad
\end{eqnarray}
This result shows that the conformal symmetry, represented by $R$ and $R_{\mu\nu}$ terms, plays no role in the critical exponents values thus confirming the universality hypothesis, at least up to the NLO order inspected in this work. In fact, this symmetry is one present in the space where the field is embedded and not in its internal one. Only an internal one would change the critical exponents values.

\section{Conclusions}\label{Conclusions}

\par We have analytically probed the effect of conformal symmetry in the NLO radiative quantum corrections to critical exponents for massless O($N$) $\lambda\phi^{4}$ scalar field theories in curved spacetime. We have applied the field-theoretic renormalization group approach in the massless BPHZ method. We have found that the conformal symmetry has not modified the NLO order critical exponents values thus keeping them at their flat spacetime values. This result is in perfect agreement with the universality hypothesis, since a given symmetry could affect the critical exponents values only if it would be present in the internal space of the fluctuating quantum field and not one in a space where the field is embedded. For our knowledge, the present work opens a new research branch, namely one in which we can investigate the influence of conformal symmetry in the critical properties of systems undergoing continuous phase transitions in curved spacetime like critical exponents in geometries subjected to different boundary conditions, finite-size scaling effects, corrections to scaling etc.      

\section*{Acknowledgements}

\par With great pleasure the authors thank the kind referee for helpful comments. HASC would like to thank CAPES (brazilian funding agency) for financial support.

\appendix*
\section{Evaluated three-loop diagrams}\label{Evaluated three-loop diagrams}

\par By evaluating analytically, through dimensional regularization in notation of Ref. \cite{PhysRevD.96.116002} and considering that $\xi = \xi(d) = [(d-2)/4(d-1)]$ for $d < 4$ and $d = 4 - \epsilon$, the needed three-loop diagrams are given by   
\begin{eqnarray}\label{gjhkjhih}
&& \parbox{12mm}{\includegraphics[scale=1.0]{fig6.eps}} = -\frac{P^{2}f^{2}}{8\epsilon}\left[1 + \frac{1}{4}\epsilon - 2J_{3}(P^{2})\epsilon\right] + \frac{Rf^{2}}{48\epsilon},\quad
\end{eqnarray}
\begin{eqnarray}\label{dfhuughfbjkdf}
&& \parbox{12mm}{\includegraphics[scale=1.0]{fig7.eps}} = \frac{P^{2}f^{3}}{6\epsilon^{2}}\left[1 + \frac{1}{2}\epsilon - 3J_{3}(P^{2})\epsilon\right] +  \frac{5Rf^{3}}{54\epsilon^{2}}\Biggl[ 1 - \frac{2}{3}\epsilon + \frac{3}{2}\tilde{i}(P^{2})\epsilon  \Biggr] - \nonumber \\&& \frac{5}{6\epsilon}R_{\mu\nu}J_{3}^{\mu\nu}(P^{2})f^{3},
\end{eqnarray}
where
\begin{eqnarray}\label{uhduhdudfgjgdhg}
J_{3}(P^{2}) = \int_{0}^{1}dx(1-x)\ln \Biggl[\frac{x(1-x)P^{2}}{\mu^{2}}\Biggr],\quad\quad
\end{eqnarray}
\begin{eqnarray}\label{uhduhufgjgdhg}
\tilde{i}(P^{2}) = \int_{0}^{1}dx(1-x)\ln x  \frac{d}{dx}\Biggl\{(1-x)\ln \Biggl[\frac{x(1-x)P^{2}}{\mu^{2}} \Biggr]\Biggr\},
\end{eqnarray}
\begin{eqnarray}\label{uhduhdudfgjgdhg}
J_{3}^{\mu\nu}(P^{2}) = \int_{0}^{1}dx\frac{x(1-x)^{2}P^{\mu}P^{\nu}}{x(1-x)P^{2}},
\end{eqnarray}
\begin{eqnarray}\label{djigjfjgi}
&&\parbox{10mm}{\includegraphics[scale=1.0]{fig26.eps}} \quad = -\frac{3P^{2}f^{3}}{16\epsilon^{2}}\left[1 + \frac{1}{4}\epsilon - 2\epsilon J_{3}(P^{2})\right],
\end{eqnarray}
where
\begin{eqnarray}
\parbox{10mm}{\includegraphics[scale=1.0]{fig26.eps}}  = 
\parbox{10mm}{\includegraphics[scale=1.0]{fig6.eps}}\,\bigg|_{R=0, -\mu^{\epsilon}f \rightarrow -\mu^{\epsilon}fc_{f}^{1}}
\end{eqnarray}
and $c_{f}^{1}$ is the one-loop order coupling constant counterterm,
\begin{eqnarray}
\parbox{10mm}{\includegraphics[scale=1.0]{fig10.eps}} = \frac{\mu^{\epsilon}f^{2}}{\epsilon} \Biggr[1 - \frac{1}{2}\epsilon - \frac{1}{2}\epsilon J(P^{2})  + \frac{R}{6\mu^{2}}\epsilon J_{R}(P^{2}) - \frac{R_{\mu\nu}P^{\mu}P^{\nu}}{3\mu^{4}}\epsilon J_{R_{\mu\nu}}(P^{2}) \Biggr],\quad
\end{eqnarray}
\begin{eqnarray}
\parbox{16mm}{\includegraphics[scale=1.0]{fig11.eps}} = -\frac{\mu^{\epsilon}f^{3}}{\epsilon^{2}} \Biggr[1 - \epsilon - \epsilon J(P^{2})  + \frac{R}{3\mu^{2}}\epsilon J_{R}(P^{2}) -  2\frac{R_{\mu\nu}P^{\mu}P^{\nu}}{3\mu^{4}}\epsilon J_{R_{\mu\nu}}(P^{2}) \Biggr],\quad \nonumber \\
\end{eqnarray}
\begin{eqnarray}
\parbox{12mm}{\includegraphics[scale=0.8]{fig21.eps}} = -\frac{\mu^{\epsilon}f^{3}}{2\epsilon^{2}} \Biggr[1 - \frac{1}{2}\epsilon - \epsilon J(P^{2})  + \frac{R}{3\mu^{2}}\epsilon J_{R}(P^{2}) -  2\frac{R_{\mu\nu}P^{\mu}P^{\nu}}{3\mu^{4}}\epsilon J_{R_{\mu\nu}}(P^{2})\Biggr],\quad\quad
\end{eqnarray}
\begin{eqnarray}
\parbox{10mm}{\includegraphics[scale=1.0]{fig25.eps}} = \frac{3\mu^{\epsilon}f^{3}}{2\epsilon^{2}} \Biggr[1 - \frac{1}{2}\epsilon - \frac{1}{2}\epsilon J(P^{2})  + \frac{R}{6\mu^{2}}\epsilon J_{R}(P^{2}) - \frac{R_{\mu\nu}P^{\mu}P^{\nu}}{3\mu^{4}}\epsilon J_{R_{\mu\nu}}(P^{2}) \Biggr],\quad\quad
\end{eqnarray}
where
\begin{eqnarray}\label{uhduhufgjg}
J(P^{2}) = \int_{0}^{1}dx \ln \left[\frac{x(1-x)P^{2}}{\mu^{2}}\right],
\end{eqnarray}
\begin{eqnarray}\label{ugujdfjgdhg}
J_{R}(P^{2}) =  \int_{0}^{1}d x\frac{x(1 - x)}{\frac{x(1 - x)P^{2}}{\mu^{2}}},
\end{eqnarray}
\begin{eqnarray}\label{ufgfghujjgdhg}
J_{R_{\mu\nu}}(P^{2}) =  \int_{0}^{1}d x\frac{x^{2}(1 - x)^{2}}{\left[\frac{x(1 - x)P^{2}}{\mu^{2}}\right]^{2}}.
\end{eqnarray}

\bibliography{apstemplate}

\end{document}